\documentclass[12pt]{article}
\usepackage[
dvips]{graphicx}
\usepackage{amsmath}
\makeatletter
\def\thesection{\arabic{section}}
\@addtoreset{equation}{section}

\def\appendix{\par
\setcounter{section}{0}
\setcounter{subsection}{0}
\def\thesection{\Alph{section}}}
\makeatother
\title{Functional Integral in terms of the Field Strength: \\
An Approach to Chiral Symmetry Breaking}
\author{Naoki~Tanimura\\
\\
Department of Physics, Kyushu University\\
Fukuoka 812-8581, JAPAN\\
{\small E-mail: {\tt naoki@higgs.phys.kyushu-u.ac.jp}}}
\date{\today}
\begin{document}
\maketitle
\begin{abstract}
The chiral symmetry breaking in the 4-dimensional QED with 
the chirally invariant four-fermion interaction is discussed by using 
a novel path integral expression in terms of the field-strength 
tensor. In the local potential approximation, we find that the 
chiral symmetry is spontaneously broken 
for any nonzero gauge and four-fermion couplings on the tree level 
of an auxiliary field $\sigma$. The present approach allows us to 
easily include higher orders of the gauge coupling so that 
the effective potential up to the sixth order is obtained. 
\end{abstract}
\section{Introduction}\label{intro}
The standard model of elementary particles has had good agreement with 
experiments, in which masses of quarks 
and leptons are generated by the Yukawa interactions with the Higgs bosons. 
There, however, have been no experimental observations of the Higgs 
boson, and moreover, there is a theoretical problem of the ``naturalness.''
Therefore models without any elementary scalars have been 
considered\cite{TC,TQC} as effective field theories to understand 
the origin of fermion masses: 
Those are generated dynamically through the interaction 
\begin{equation}
	{g^{2}\over2}\left[(\overline\psi\psi)^{2}
	+(\overline{\psi}i\gamma_{5}\psi)^{2}\right],
\end{equation}
which is suppressed by inverse powers of a high energy scale $\Lambda$, 
at energies much below it, that is, $g^{2}\Lambda^{2}\sim O(1)$. 
It is called the Nambu--Jona-Lasinio (NJL) model\cite{NJL} and has been 
generalized to the one with the gauge interaction\cite{BLL}: 
\begin{equation}
	{\mathcal{L}}=-{1\over4}F_{\mu\nu}F^{\mu\nu}
	+\overline\psi\,i\gamma_{\mu}(\partial^{\mu}-ieA^{\mu})\psi
	+{g^{2}\over2}\left[(\overline\psi\psi)^{2}
	+(\overline{\psi}i\gamma_{5}\psi)^{2}\right].
	\label{LQED}
\end{equation}
Although, in this model, nonperturbative approaches such as the 
Schwinger--Dyson equation with the ladder approximation\cite{KMY,ASTW}
or renormalization group\cite{AMSTT} give a nontrivial phase structure 
for chiral symmetry in a 
plane of gauge and four-fermi couplings, they suffer from gauge 
dependence. In the perturbative framework the inversion method\cite{FKYSOI} 
does give a gauge-independent result\cite{Ko}, 
higher order calculations are tough tasks. 

If we employ the perturbative effective action approach to the problem 
of chiral symmetry breaking($\chi$SB), there is some room for improvement or 
simplification of the calculation from two viewpoints: (i) 
In view of the nonperturbative renormalization group 
approach\cite{AMSTT}, the critical behavior 
is successfully described in the local potential approximation: 
taking the lowest order of derivative expansion in the effective 
action. We, therefore, expect 
that it could also be good in the perturbative calculation. 
%
(ii) After integrating with respect to matter (fermionic) fields the 
result must be gauge invariant, that is, be written as a functional of 
the field-strength tensor $F_{\mu\nu}$. Therefore it is preferable to 
rewrite the functional integral of the gauge potential $A_{\mu}$ into 
that of $F_{\mu\nu}$. This could be a great help to the gauge-invariant 
calculation. 

Indeed in the lattice gauge 
theory, a change of variables from gauge potentials to field strengths 
has been given\cite{Wi}, whose recipe, however, is specialized to the 
lattice. In the continuum theory, the field-strength formulation has 
already been proposed\cite{MD}. It is, however, based on the special 
choice of gauge---the coordinate gauge $x_{\mu}A^{\mu}=0$, which 
cannot fix the gauge at the origin and moreover yields complicated 
fermionic currents with line integrals in the lagrangian. 

In this article we give a perturbative and gauge-independent method 
for calculating the dynamical fermion mass under the local potential 
approximation. Inclusion of higher orders of the gauge coupling 
constant is simpler in this method. 
We use the functional integration not as a method simply 
reproducing perturbative diagrams systematically, but as 
``integration'' with 
nice features: Changing variables and introducing auxiliary variables 
can easily be made. 
We first construct a Euclidean path 
integral expression in terms of the gauge field strength for the gauge 
sector with a conserved source $J_{\mu}$: 
We start from the canonical 
formalism for quantization and find a suitable change of variables, 
which is the contents of Sec.~\ref{FSPI}. 
Second, in Sec.~\ref{CSB} the fermionic partition function, minimally 
coupled to the gauge potential, in the local 
potential approximation is obtained by introducing auxiliary fields 
and by utilizing the Fock-Schwinger 
proper-time method\cite{FS,IKT}. 
At last combining the above two partition functions and integrating 
with respect to the gauge field strength to obtain an effective 
potential of the 
dynamical fermion mass. The final Sec.~\ref{DISC} is devoted to a 
discussion. 

\section{Functional Integral in terms of the Field Strength}\label{FSPI}

In this section we construct a Euclidean path integral expression 
in 4-dimensional abelian gauge theory coupled to a 
conserved current $J_{\mu}$, based on the canonical 
formalism\hspace{0pt}\cite{KT}, 
\begin{equation}
	{\mathcal{L}}=-{1\over4}F_{\mu\nu}F^{\mu\nu}+J_{\mu}A^{\mu}.
\end{equation}

In order to show the 
gauge independence manifestly we fix the gauge in terms of an arbitrary 
real function $\phi_{\mu}(x)$ satisfying\cite{KT,DEMST} 
\begin{equation}
  \partial_\mu\phi^\mu(x)=\delta^4(x);\label{phi}
\end{equation}
the gauge fixing condition is\cite{KT} 
\begin{equation}
	\int\!\!d^{4}y\;\phi_{\mu}(x-y)A^{\mu}(y)=0, 
	\label{gfc}
\end{equation}
which is satisfied by 
\begin{equation}
A^\mu(x;\phi)=A^{\mu}(x)-\partial^\mu_x\int\!\!d^4y\,
\phi_\nu(x-y)A^\nu(y),\label{physA}
\end{equation}
for an arbitrary $A^{\mu}(x)$. 
%
%
%
Here we take a $\phi_{\mu}(x)$ whose support is 
3-dimensional and spacelike: 
\begin{equation}
\phi^{\mu}(x)=(0,f^{i}({\boldsymbol{x}})\delta(x_0)),\hspace{30pt}
\widetilde{\phi}^{\mu}(p)=(0,\widetilde{f}^{i}({\boldsymbol{p}}))
\label{phi1}
\end{equation}
with 
\begin{equation}
	{\nabla}^{i}f^{i}(x)=\delta^{3}({\boldsymbol{x}}),
	\hspace{30pt}{p}^{i}\widetilde f^{i}({\boldsymbol{p}})=-i.
	\label{fcond}
\end{equation}
If it were not for this restriction, $A^\mu(x;\phi)$ 
in Eq.~(\ref{physA}) is nonlocal in time, so that we 
cannot follow the canonical procedure. 
(Here and hereafter repeated indices $i$, $j$, $k$, etc., imply the 
summation over 1 to 3.)
The relation (\ref{physA}) turns into 
\begin{eqnarray}
A^{0}({\boldsymbol{x}};\phi)&=&A^{0}({\boldsymbol{x}})
+\widetilde{f}^j(-i{\boldsymbol{\nabla}})
\dot{A}^{j}({\boldsymbol{x}}),\\
A^i({\boldsymbol{x}};\phi)&=&A^i({\boldsymbol{x}})
-{\nabla}^i 
\widetilde{f}^j(-i{\boldsymbol{\nabla}})A^j({\boldsymbol{x}}),\label{b9}
\end{eqnarray}
where 
\begin{equation}
	\widetilde{f}^{j}(-i{\boldsymbol{\nabla}})A^{j}({\boldsymbol{x}})\equiv
	\int\!\!{d^{3}{\boldsymbol{p}}\over(2\pi)^{3}}
	{\mathrm{e}}^{i{\boldsymbol{p}}\cdot{\boldsymbol{x}}}
	\widetilde{f}^{j}({\boldsymbol{p}})\widetilde{A}^{j}({\boldsymbol{p}})
	=\int\!\!d^{3}y\,f^{j}({\boldsymbol{x}}-{\boldsymbol{y}})A^{j}({\boldsymbol{y}})
	\label{ABB}
\end{equation}
and 
the time argument has been omitted. 
Hereafter we use an abbreviation like Eq.~(\ref{ABB}) for simplicity; 
all the functions of $-i{\boldsymbol{\nabla}}$ including 
$1/\vert{\boldsymbol{\nabla}}\vert$ and $1/{\boldsymbol{\nabla}}^{2}$ should be 
understood in the momentum space representation. 

The gauge-fixed variable $A^{i}({\boldsymbol{x}};\phi)$ must have two 
degrees of freedom out of three $A^{i}({\boldsymbol{x}})$'s, which can be 
singled out by considering the norm of the functional space
\begin{equation}
	\int\!\!d^{3}x\,\delta A^{i}({\boldsymbol{x}};\phi)
	\delta A^{i}({\boldsymbol{x}};\phi)=\int\!\!d^{3}x\,
	\delta A^{j}({\boldsymbol{x}})M^{jk}(-i{\boldsymbol{\nabla}})
	\delta A^{k}({\boldsymbol{x}}),
	\label{norm}
\end{equation}
where 
\begin{equation}
	M^{jk}({\boldsymbol{p}})\equiv\delta^{jk}
	+i\widetilde{f}^{j*}({\boldsymbol{p}})p^{k}
	-ip^{j}\widetilde{f}^{k}({\boldsymbol{p}})
	+{\boldsymbol{p}}^{2}
	\widetilde{f}^{j*}({\boldsymbol{p}})\widetilde{f}^{k}({\boldsymbol{p}}).
\end{equation}
This matrix can be diagonalized as 
\begin{equation}
	n^{i}_{(\alpha)}({\boldsymbol{p}})M^{ij}({\boldsymbol{p}})
	n^{j*}_{(\beta)}({\boldsymbol{p}})=\left(
	\begin{array}{ccc}
		1&&\\&{\boldsymbol{p}}^{2}\vert
		\widetilde{\boldsymbol{f}}({\boldsymbol{p}})\vert^{2}&\\
		&&0
	\end{array}\right)_{\alpha\beta},
	\label{Mdiag}
\end{equation}
where $n_{(\alpha)}^{k}$'s are given by 
\begin{eqnarray}
n_{(1)}^k({\boldsymbol{p}})&\equiv&\epsilon^{klm}
n_{(2)}^{l*}({\boldsymbol{p}})\,n_{(3)}^{m*}({\boldsymbol{p}}),\nonumber\\
n_{(2)}^k({\boldsymbol{p}})&\equiv&
\left[ip^k+{\boldsymbol{p}}^2\widetilde f^k({\boldsymbol{p}})\right]\Big/
\sqrt{{\boldsymbol{p}}^2({\boldsymbol{p}}^2\vert\widetilde{\boldsymbol f}({\boldsymbol{p}})\vert^2-1)},
\label{ns}\\
n_{(3)}^k({\boldsymbol{p}})&\equiv&ip^k/\vert{\boldsymbol{p}}\vert.\nonumber
\end{eqnarray}
and form an orthonormal base obeying 
\begin{equation}
\sum_{\alpha=1}^3n_{(\alpha)}^{j*}({\boldsymbol{p}})n_{(\alpha)}^k({\boldsymbol{p}})
=\delta^{jk}, \hspace{20pt}n_{(\alpha)}^{k*}({\boldsymbol{p}})
=n_{(\alpha)}^k(-{\boldsymbol{p}}).\label{comple}
\end{equation}

In view of Eqs.~(\ref{norm}) and (\ref{Mdiag}), genuine physical 
variables are given as 
\begin{equation}
A^{(\alpha)}({\boldsymbol{x}})\equiv 
n_{(\alpha)}^{k}(-i{\boldsymbol \nabla})A^{k}({\boldsymbol{x}})
=n_{(\alpha)}^{k}(-i{\boldsymbol \nabla})A^{k}({\boldsymbol{x}};\phi),
\hspace{10pt}(\alpha=1,2).
\end{equation}
With these variables the action is 
\begin{eqnarray}
S&=&\int\!\!d^{4}x\,\left\{-{1\over4}
\left[\partial_{\mu}A_{\nu}(x;\phi)-\partial_{\nu}A_{\mu}(x;\phi)
\right]^{2}
+J_{\mu}(x)A^{\mu}(x;\phi)\right\}\nonumber\\
&=&\int\!\!d^{4}x\,\Biggl\{{1\over2}\sum_{\alpha=1}^{2}
A^{(\alpha)}(x){\boldsymbol{\nabla}}^{2}A^{(\alpha)}(x)
+{1\over2}\left[\dot{A}^{(1)}(x)\right]^{2}
\nonumber\\
&&{}-{1\over2}{\dot{A}^{(2)}}(x)\!{\boldsymbol{\nabla}}^{2}
\vert\widetilde{\boldsymbol f}(-i{\boldsymbol{\nabla}})\vert^{2}
{\dot{A}^{(2)}}(x)
+{1\over2}\left[{\boldsymbol{\nabla}} A^{0}(x;\phi)\right]^{2}
\nonumber\\
&&{}-A^{0}(x;\phi)   
\sqrt{{\boldsymbol{\nabla}}^2({\boldsymbol{\nabla}}^2
\vert\widetilde{\boldsymbol f}(-i{\boldsymbol{\nabla}})\vert^2+1)}
{\dot{A}^{(2)}}(x) 
\label{b22}\\
&&{}+J_{0}(x)A^{0}(x;\phi)
-{\boldsymbol J}(x)\!\cdot\!\left[{\boldsymbol A}(x;\phi)
\right]\Biggr\}, \nonumber 
\end{eqnarray}
where $\left[{\boldsymbol A}(x;\phi)\right]$ in the last line is given by 
\begin{eqnarray}
A^i({\boldsymbol{x}};\phi)
&=&n_{(1)}^{i*}(-i{\boldsymbol{\nabla}})A^{(1)}({\boldsymbol{x}})
\label{aphi3}\\
&&{}+\left[n_{(2)}^{i*}(-i{\boldsymbol{\nabla}})
+n_{(3)}^{i*}(-i{\boldsymbol{\nabla}})
\sqrt{-{\boldsymbol{\nabla}}^{2}
\vert{\boldsymbol{f}}(-i{\boldsymbol{\nabla}})\vert^{2}-1}
\right]A^{(2)}({\boldsymbol{x}}).
\nonumber
\end{eqnarray}

The Hamiltonian is written as 
\begin{eqnarray}
H(t)&=&\int\!\!d^{3}x\,\Biggl\{{1\over2}\sum_{\alpha=1}^{2}
\left(\left[{\mit\Pi}^{(\alpha)}({\boldsymbol{x}})\right]^{2}
+\left[{\boldsymbol{\nabla}}
A^{(\alpha)}({\boldsymbol{x}})\right]^{2}\right)
\nonumber\\
&&{}+J_{0}(x){\sqrt{{\boldsymbol{\nabla}}^2({\boldsymbol{\nabla}}^2
\vert\widetilde{\boldsymbol{f}}(-i{\boldsymbol{\nabla}})\vert^2+1)}
\over{\boldsymbol{\nabla}}^{2}}{\mit\Pi}^{(2)}({\boldsymbol{x}})
\\
&&{}+{1\over2}J_{0}(x)
\vert\widetilde{\boldsymbol{f}}(-i{\boldsymbol{\nabla}})\vert^{2}J_{0}(x)
+{\boldsymbol{J}}(x)\!\cdot\!\left[{\boldsymbol{A}}({\boldsymbol{x}};\phi)\right]
\Biggr\}\nonumber
\end{eqnarray}
in terms of four dynamical variables, 
two $A^{(\alpha)}({\boldsymbol{x}})$'s and 
their canonical conjugate momenta 
\begin{eqnarray}
	{\mit\Pi}^{(1)}({\boldsymbol x})&=&
	\dot{A}^{(1)}({\boldsymbol x}),
	\label{pi1}  \\
	{\mit\Pi}^{(2)}({\boldsymbol x})&=&
	\dot{A}^{(2)}({\boldsymbol x})
	-{\sqrt{{\boldsymbol{\nabla}}^2({\boldsymbol{\nabla}}^2
	\vert\widetilde{\boldsymbol f}(-i{\boldsymbol{\nabla}})\vert^2+1)}
	\over{\boldsymbol{\nabla}}^{2}}J_{0}(t,{\boldsymbol x}),
	\label{pi2}
\end{eqnarray}
where $A^{0}({\boldsymbol{x}};\phi)$ has been eliminated by 
\begin{equation}
	{\sqrt{{\boldsymbol{\nabla}}^2({\boldsymbol{\nabla}}^2
	\vert\widetilde{\boldsymbol f}(-i{\boldsymbol{\nabla}})\vert^2+1)}
	\over{\boldsymbol{\nabla}}^{2}\vert\widetilde{\boldsymbol f}(-i{\boldsymbol{\nabla}})\vert^2}
	{\mit \Pi}^{(2)}({\boldsymbol x})
	+{1\over\vert\widetilde{\boldsymbol f}(-i{\boldsymbol{\nabla}})\vert^2}
	A^{0}({\boldsymbol x};\phi)
	+J_{0}(t,{\boldsymbol x})=0.
	\label{constr}
\end{equation}
The field strengths are given as
\begin{eqnarray}
	E^{i}&\equiv&F^{i0}\nonumber\\
	&=&-n_{(1)}^{i*}(-i{\boldsymbol{\nabla}}){\mit\Pi}^{(1)}
	-n_{(2)}^{i*}(-i{\boldsymbol{\nabla}}){\mit\Pi}^{(2)}
	+\widetilde{f}^{i*}(-i{\boldsymbol{\nabla}})J_{0},\label{E}\\
	B^{i}&\equiv&-{1\over2}\epsilon^{ijk}F_{jk}\nonumber\\
	&=&\vert{\boldsymbol{\nabla}}\vert n_{(1)}^{i}(-i{\boldsymbol{\nabla}})A^{(2)}
	-\vert{\boldsymbol{\nabla}}\vert n_{(2)}^{i}(-i{\boldsymbol{\nabla}})A^{(1)}.
	\label{B}
\end{eqnarray}
Following the standard canonical procedure we obtain the 
path integral representation of the partition function 
$
Z_{T}[J]={\mathrm{Tr}}({\mathrm{e}}^{-TH})
$\cite{KT}, 
\begin{eqnarray}
Z_{T}[J]&=&\int\!\!{\mathcal{D}}{\mit\Pi}^{(\alpha)}
{\mathcal{D}}A^{(\alpha)}\exp\Biggl[\int\!\!d^{4}x_{_{E}}
\biggl\{i\sum_{\alpha=1}^{2}
{\mit \Pi}^{(\alpha)}(\tau,{\boldsymbol{x}}) 
\dot{A}^{(\alpha)}(\tau,{\boldsymbol{x}})  
\nonumber\\
&&-{1\over2}\sum_{\alpha=1}^{2}
\left(\left[{\mit\Pi}^{(\alpha)}(\tau,{\boldsymbol{x}})\right]^{2}
+\left[{\boldsymbol{\nabla}}A^{(\alpha)}(\tau,{\boldsymbol{x}})
\right]^{2}\right)
\nonumber\\
&&+iJ_{4}(\tau,{\boldsymbol{x}})
{\sqrt{{\boldsymbol{\nabla}}^2({\boldsymbol{\nabla}}^2
\vert\widetilde{\boldsymbol{f}}(-i{\boldsymbol{\nabla}})\vert^2+1)}
\over{\boldsymbol{\nabla}}^{2}}{\mit\Pi}^{(2)}(\tau,{\boldsymbol{x}})
\nonumber\\
&&+{1\over2}J_{4}(\tau,{\boldsymbol{x}})
\vert\widetilde{\boldsymbol{f}}(-i{\boldsymbol{\nabla}})\vert^{2}
J_{4}(\tau,{\boldsymbol{x}})
-{\boldsymbol{J}}(\tau,{\boldsymbol{x}})\!\cdot\!
[{\boldsymbol{A}}(\tau,{\boldsymbol{x}};\phi)]\biggr\}\Biggr],\label{Z}
\end{eqnarray}
where $\bigl[{\boldsymbol{A}}(\tau,{\boldsymbol{x}};\phi)\big]$ is given by 
Eq.~(\ref{aphi3}) with ${\boldsymbol{A}}({\boldsymbol{x}};\phi)\rightarrow
{\boldsymbol{A}}(\tau,{\boldsymbol{x}};\phi)$, 
\begin{equation}
	\int\!\!d^{4}x_{_{E}}\equiv\int_{0}^{T}\!\!d\tau 
	\int\!\!d^{3}x,\quad J_{4}\equiv iJ_{0},
\end{equation}
and the periodic boundary condition 
$A^{(\alpha)}(T,{\boldsymbol{x}};\phi)=A^{(\alpha)}(0,{\boldsymbol{x}};\phi)$ 
should be understood. 

Our purpose is to rewrite the expression (\ref{Z}) of four variables 
${\mit\Pi}^{\alpha}$ and $A^{\alpha}$ into that of six variables $E^{i}$ 
and $B^{i}$. In view of Eq.~(\ref{comple}) quantities which are 
proportional to $n_{(3)}^{i*}(-i{\boldsymbol{\nabla}})$ 
or $n_{(3)}^{i}(-i{\boldsymbol{\nabla}})$ 
are missing in their expressions (\ref{E}) and 
(\ref{B}). 
To this end, we introduce $\varepsilon$ and $\beta$ 
with the aid of the delta functions in the functional measure, 
\begin{eqnarray}
	E^{i}&=&-n_{(1)}^{i*}(\!-i{\boldsymbol{\nabla}}\!){\mit\Pi}^{(1)}
	-n_{(2)}^{i*}(\!-i{\boldsymbol{\nabla}}\!){\mit\Pi}^{(2)}
	+n_{(3)}^{i*}(\!-i{\boldsymbol{\nabla}}\!)\varepsilon
	-i\widetilde{f}^{i*}(\!-i{\boldsymbol{\nabla}}\!)J_{4},\label{IE}\\
	B^{i}&=&\vert{\boldsymbol{\nabla}}\vert
	n_{(1)}^{i}(\!-i{\boldsymbol{\nabla}}\!)A^{(2)}
	-\vert{\boldsymbol{\nabla}}\vert
	n_{(2)}^{i}(\!-i{\boldsymbol{\nabla}}\!)A^{(1)}
	+n_{(3)}^{i}(\!-i{\boldsymbol{\nabla}}\!)\beta,
	\label{IB}
\end{eqnarray}
or equivalently 
\begin{eqnarray}
	\varepsilon & = & n_{(3)}^{i}(-i{\boldsymbol{\nabla}})
	[{E}^{i}+i\widetilde{f}^{i*}(-i{\boldsymbol{\nabla}})J_{4}]
	={1\over\vert{\boldsymbol{\nabla}}\vert}({\nabla}^{i}{E}^{i}-iJ_{4}),
	\\
	\beta & = & n_{(3)}^{i*}(-i{\boldsymbol{\nabla}})B^{i}
	={1\over\vert{\boldsymbol{\nabla}}\vert}{\nabla}^{i}B^{i}.
\end{eqnarray}
The norm of the functional space is given by 
\begin{eqnarray}
	\int\!\!d^{3}x\,\delta E^{i}\delta E^{i}&=&
	\int\!\!d^{3}x\,(\delta{\mit\Pi}^{(1)}\delta{\mit\Pi}^{(1)}
	+\delta{\mit\Pi}^{(2)}\delta{\mit\Pi}^{(2)}
	+\delta\varepsilon\delta\varepsilon),
	\\
	\int\!\!d^{3}x\,\delta B^{i}\delta B^{i}&=&
	\int\!\!d^{3}x\,(-\delta{A}^{(1)}{\boldsymbol{\nabla}}^{2}\delta{A}^{(1)}
	-\delta{A}^{(2)}{\boldsymbol{\nabla}}^{2}\delta{A}^{(2)}
	+\delta\beta\delta\beta),
\end{eqnarray}
so that 
\begin{equation}
	{\mathcal{D}}{\mit\Pi}^{(\alpha)}{\mathcal{D}}A^{(\alpha)}{\mathcal{D}}\varepsilon
	{\mathcal{D}}\beta\prod_{x}\delta(\varepsilon)\delta(\beta)
	={\mathcal{D}}{E}^{i}{\mathcal{D}}B^{i}
	\prod_{x}\delta({\nabla}^{i}{E}^{i}-iJ_{4})
	\delta({\nabla}^{i}B^{i}).
	\label{Jacobi}
\end{equation}
We then arrived at the desired result 
\begin{eqnarray}
Z_{T}[J]&=&\int\!\!{\mathcal{D}}E^{i}{\mathcal{D}}B^{i}
	\prod_{x}\delta({\nabla}^{i}E^{i}-iJ_{4})
	\delta({\nabla}^{i}B^{i})
	\nonumber\\
	&&{}\times
	\exp\Biggl[\int\!\!d^{4}x_{_{E}}
	\biggl\{iE^{i}(\tau,{\boldsymbol{x}})\epsilon^{ijk}
	{{{\nabla}}^{j}\over{\boldsymbol{\nabla}}^{2}}
	\dot{B}^{k}(\tau,{\boldsymbol{x}})
	-{1\over2}\left(\left[E^{i}(\tau,{\boldsymbol{x}})\right]^{2}
	+\left[B^{i}(\tau,{\boldsymbol{x}})\right]^{2}\right)
	\nonumber\\
	&&{}+J^{i}(\tau,{\boldsymbol{x}})\epsilon^{ijk}
	{{{\nabla}}^{j}\over{\boldsymbol{\nabla}}^{2}}
	B^{k}(\tau,{\boldsymbol{x}})
	\biggr\}\Biggr],
	\label{ZF}
\end{eqnarray}
which is apparently free from the choice of $f^{i}({\boldsymbol{x}})$ as 
expected. 
%

To carry out the integration in the next section, 
it is convenient to introduce ${\overline{E}}^{i}$ as 
\begin{equation}
	{\overline{E}}^{i}\equiv 
	E^{i}-i{\nabla^{i}\over{\boldsymbol{\nabla}}^{2}}J_{4},
\end{equation}
so that 
\begin{equation}
	{{\nabla}}^{i}{\overline E}^{i}={{\nabla}}^{i}E^{i}-iJ_{4},
	\label{EB}
\end{equation}
giving 
\begin{eqnarray}
Z_{T}[J]&=&\int\!\!{\mathcal{D}}{\overline E}^{i}{\mathcal{D}}B^{i}
	\prod_{x}\delta({\nabla}^{i}{\overline E}^{i})
	\delta({\nabla}^{i}B^{i})
	\exp\Biggl[\int\!\!d^{4}x_{_{E}}
	\biggl\{i{\overline E}^{i}(\tau,{\boldsymbol{x}})\epsilon^{ijk}
	{{{\nabla}}^{j}\over{\boldsymbol{\nabla}}^{2}}
	\dot{B}^{k}(\tau,{\boldsymbol{x}})
	\nonumber\\
	&&{}-{1\over2}\left(\left[
	{\overline E}^{i}(\tau,{\boldsymbol{x}})\right]^{2}
	+\left[B^{i}(\tau,{\boldsymbol{x}})\right]^{2}\right)
	-{1\over2}J_{4}(\tau,{\boldsymbol{x}})
	{1\over{\boldsymbol{\nabla}}^{2}}J_{4}(\tau,{\boldsymbol{x}})
	\nonumber\\
	&&{}+J^{i}(\tau,{\boldsymbol{x}})\epsilon^{ijk}
	{{{\nabla}}^{j}\over{\boldsymbol{\nabla}}^{2}}
	B^{k}(\tau,{\boldsymbol{x}})
	\biggr\}\Biggr].
	\label{ZC}
\end{eqnarray}
Further introducing an auxiliary field $\rho$: 
\begin{equation}
	1=\int\!\!{\mathcal{D}}\rho\, 
	\exp\left[-{1\over2}\int\!\!d^{4}x_{_{E}}
	\left(\rho+{1\over\vert{\boldsymbol{\nabla}}\vert}J_{4}\right)^{2}\right],
\end{equation}
so as to cancel the $(J_{4})^{2}$ term in Eq.~(\ref{ZC}), 
decomposing ${\overline E}^{i}$ and $B^{i}$ as
\footnote{Any choice for the complete orthonormal base 
$\{n_{(\alpha)}^{i}(-i{\boldsymbol{\nabla}})\}$ can be taken. Here, we only 
assume the same $n_{(3)}^{i}(-i{\boldsymbol{\nabla}})$ as Eq.~(\ref{ns}).}
\begin{equation}
	{\overline E}^{i}
	=\sum_{\alpha=1}^{3}n_{(\alpha)}^{i*}(-i{\boldsymbol{\nabla}})
	\varepsilon_{\alpha},\hspace{20pt}
	B^{i}=\sum_{\alpha=1}^{3}n_{(\alpha)}^{i}(-i{\boldsymbol{\nabla}})
	\beta_{\alpha},
\end{equation}
and integrating with respect to $\varepsilon_{\alpha}$'s and $\beta_{3}$, 
we obtain 
\begin{eqnarray}
Z_{T}[J]&=&\int\!\!{\mathcal{D}}{\beta}_{\alpha}{\mathcal{D}}\rho
[{\mathrm{Det}}\vert{\boldsymbol{\nabla}}\vert]^{-2}
\exp\Biggl[\int\!\!d^{4}x_{_{E}}\biggl\{
-{1\over2}\sum_{\alpha=1}^{2}
\left(\left[{1\over\vert{\boldsymbol{\nabla}}\vert}
\dot{\beta}_{\alpha}(\tau,{\boldsymbol{x}})\right]^{2}
+\beta_{\alpha}(\tau,{\boldsymbol{x}})^{2}\right)
\nonumber\\
&&-{1\over2}\rho(\tau,{\boldsymbol{x}})^{2}
-J_{4}(\tau,{\boldsymbol{x}})
{1\over\vert{\boldsymbol{\nabla}}\vert}\rho(\tau,{\boldsymbol{x}})\nonumber\\
&&-J_{i}(\tau,{\boldsymbol{x}})
{1\over\vert{\boldsymbol{\nabla}}\vert}
\left[n_{(2)}^{i*}(-i{\boldsymbol{\nabla}})\beta_{1}(\tau,{\boldsymbol{x}})
-n_{(1)}^{i*}(-i{\boldsymbol{\nabla}})\beta_{2}(\tau,{\boldsymbol{x}})\right]
\biggr\}\Biggr].\label{Z3}
\end{eqnarray}
This form with further changes of variables is used in the next section, 
where we regard the coefficients of the sources in Eq.~(\ref{Z3}) as 
gauge potentials: 
\begin{eqnarray}
	{\overline A}_{4}&\equiv&{1\over\vert{\boldsymbol{\nabla}}\vert}\rho,
	\label{A4B}  \\
	{\overline A}_{i}&\equiv&{1\over\vert{\boldsymbol{\nabla}}\vert}
	\left[n_{(2)}^{i*}(-i{\boldsymbol{\nabla}})\beta_{1}
	-n_{(1)}^{i*}(-i{\boldsymbol{\nabla}})\beta_{2}\right].
	\label{AiB}
\end{eqnarray}
Therefore field strengths are given as 
\begin{eqnarray}
	{\overline F}_{i4}&\equiv&{{\nabla}}_{i}{\overline A}_{4}
	-\dot{\overline A}_{i}
	\nonumber  \\
	&=&n_{(1)}^{i*}(-i{\boldsymbol{\nabla}})
	{1\over\vert{\boldsymbol{\nabla}}\vert}\dot{\beta}_{2}
	-n_{(2)}^{i*}(-i{\boldsymbol{\nabla}})
	{1\over\vert{\boldsymbol{\nabla}}\vert}\dot{\beta}_{1}
	-n_{(3)}^{i*}(-i{\boldsymbol{\nabla}})\rho,
	\label{Fi4B}  \\
	{\overline F}_{ij}&\equiv&{{\nabla}}_{i}{\overline A}_{j}
	-{{\nabla}}_{j}{\overline A}_{i}
	\nonumber  \\
	&=&\epsilon_{ijk}\left[n_{(1)}^{k}(-i{\boldsymbol{\nabla}})\beta_{1}
	 +n_{(2)}^{k}(-i{\boldsymbol{\nabla}})\beta_{2}\right],
	\label{FijB}
\end{eqnarray}
so that 
\begin{equation}
	\int\!\!d^{4}x_{_{E}}{\overline F}_{\mu\nu}{\overline F}_{\mu\nu} 
	=2\int\!\!d^{4}x_{_{E}}\left\{\sum_{\alpha=1}^{2}
	\left(\left[{1\over\vert{\boldsymbol{\nabla}}\vert}
	\dot{\beta}_{\alpha}\right]^{2}
	+\beta_{\alpha}^{2}\right)+\rho^{2}\right\}.
	\label{FF}
\end{equation}

\section{$\chi$SB in QED$_{4}$ with the chirally invariant 
four-fermion interaction}\label{CSB}

In this section, we consider fermionic system coupled minimally 
to the ``gauge potentials'' (\ref{A4B}) and (\ref{AiB}) with 
the chirally invariant 
four-fermion interaction. The partition function is 
\begin{equation}
	Z[{\overline{A}}]=\int\!\!{\mathcal{D}}\psi{\mathcal{D}}{\overline\psi}
	\exp\left[\int\!\!d^{4}x_{_{E}}\left\{-{\overline\psi}\gamma_{\mu}
	(\partial_{\mu}-i{\overline{A}}_{\mu})\psi
	+{g^{2}\over2}\left[({\overline\psi}\psi)^{2}
	+({\overline\psi}i\gamma_{5}\psi)^{2}\right]\right\}\right],
	\label{ZA}
\end{equation}
where the gauge coupling constant has been absorbed into 
${\overline A}_{\mu}$. 
Our scenario is as follows: introduce auxiliary fields to cancel 
the four-fermion interaction, and then 
integrate with respect to the fermionic fields and finally 
the gauge field strengths 
with the aid of the representation in the previous section. 
The result 
is a tree potential of the auxiliary field, with 
which we examine the dynamical mass generation of fermions. 

After introducing auxiliary fields $\sigma$ and $\pi$, as usual, 
and integrating with respect to $\psi$ and ${\overline\psi}$, we have 
\begin{equation}
	Z[{\overline{A}}]=\int\!\!{\mathcal{D}}\sigma{\mathcal{D}}\pi\exp\left[
	-\int\!\!d^{4}x_{_{E}}{\sigma^{2}+\pi^{2}\over2g^{2}}
	+\ln{\mathrm{Det}}\left[\gamma_{\mu}(\partial_{\mu}
	-i{\overline{A}}_{\mu})+\sigma+i\gamma_{5}\pi\right]\right].
	\label{ZA1}
\end{equation}
Shifting as $\sigma\rightarrow m+\sigma'$ and 
$\pi\rightarrow\pi'$ and ignoring $\sigma'$ and $\pi'$, we obtain 
the tree level potential of $\sigma$($m$):
\begin{equation}
	Z[{\overline{A}}]_{0}=
	\exp\left[-\int\!\!d^{4}x_{_{E}}{m^{2}\over2g^{2}}
	+\ln{\mathrm{Det}}\left[\gamma_{\mu}(\partial_{\mu}
	-i{\overline{A}}_{\mu})+m\right]\right].
	\label{ZA2}
\end{equation}
The exponent of Eq.~(\ref{ZA2}) must be a functional of the field 
strength ${\overline{F}}_{\mu\nu}$ rather than 
${\overline{A}_{\mu}}$ as far as a regularization preserves the gauge 
invariance. 

We here employ the local potential approximation: We adopt the 
lowest order of derivative expansion, that is, discard any terms with 
differentials like $F_{\mu\nu}\framebox(7,7){}_{_{E}}F_{\mu\nu}$, to 
obtain a polynomial of $F_{\mu\nu}$. 
This approximation seems to be valid, since we are 
interested only in a low energy phenomena, the chiral symmetry 
breaking, where contributions from large $p_{\mu}$ should be 
much less important. 

The functional form of the effective action under this approximation can be 
obtained nonperturbatively 
by the Fock--Schwinger's proper time method\cite{FS,IKT}: 
\begin{equation}
	Z[{\overline{A}}]_{0}=
	\exp\left[-\int\!\!d^{^{D}}\!\!x_{_{E}}\left\{{m^{2}\over2g^{2}}
	+{1\over2(2\pi)^{{D\over2}}}\lim_{s\rightarrow0}
	\int_{0}^{\infty}\!\!d\tau\,
	\tau^{s-{D\over2}-1}{\mathrm{e}}^{-\tau m^{2}}G(\tau F)
	\right\}\right],
	\label{ZA3}
\end{equation}
where 
\begin{eqnarray}
	G(F) & = & F_{+}F_{-}\coth(F_{+})\coth(F_{-}),
	\label{G}  \\
	F_{\pm} & = & {1\over2}
	\left(\sqrt{F_{\mu\nu}F_{\mu\nu}
	+F_{\mu\nu}{\widetilde{F}}_{\mu\nu}\over2}
	\pm\sqrt{F_{\mu\nu}F_{\mu\nu}
	-F_{\mu\nu}{\widetilde{F}}_{\mu\nu}\over2}\right).
	\label{Fpm}
\end{eqnarray}
To evaluate the $\tau$-integration in Eq.~(\ref{ZA3}), we expand $G(F)$ as
\begin{eqnarray}
	G(F)&=&1+{1\over3}(F_{+}^{2}+F_{-}^{2})
	-{1\over45}\left[(F_{+}^{2}+F_{-}^{2})^{2}
	-7F_{+}^{2}F_{-}^{2}\right]
	\nonumber\\
	&&+{1\over945}\left[2(F_{+}^{2}+F_{-}^{2})^{3}
	-13F_{+}^{2}F_{-}^{2}(F_{+}^{2}+F_{-}^{2})\right]+O(F^{8}).
	\label{GE}
\end{eqnarray}
We need some regularization: 
In order to reproduce the NJL 
result in the limit $e\rightarrow0$, we need an ultraviolet 
cutoff $\Lambda$ which is introduced by a modification of the range of 
the $\tau$-integration to $[1/\Lambda^{2},\infty)$. 
However, this proper-time cutoff breaks the gauge invariance in the 
same way as the momentum-space cutoff, contrary to the dimensional 
regularization ($D=4-2\epsilon$). 
To overcome this difficulty, we employ both regularizations at the 
same time. We use the cutoff for the zeroth order in the gauge coupling, 
since it has nothing to do with gauge fields. While, for higher orders, 
the dimensional regularization is used\footnote{If we adopt the 
dimensional regularization from the zeroth order we have a different 
critical coupling due to the lack of a quadratic term of the 
renormalization scale $\mu$ in the effective action. Compare two 
expressions of the zeroth order effective action, 
Eq.~(\ref{vo}) by the proper-time cutoff after 
expanding in terms of $m^{2}/\Lambda^{2}$, 
$$V_{0}(m)={m^{2}\over2g^{2}}
+{m^{4}\over16\pi^{2}}\left[{\Lambda^{4}\over m^{4}}
-2{\Lambda^{2}\over m^{2}}-\gamma+{3\over2}
+\ln{\Lambda^{2}\over m^{2}}\right],$$
and one by the dimensional regularization, 
$$V_{0}(m)=
{m^{2}\over2g^{2}}
+{m^{4}\over16\pi^{2}}\left[{1\over\epsilon}
+\ln2\pi-\gamma+{3\over2}
+\ln{\mu^{2}\over m^{2}}\right].$$}. 
The result is 
\begin{eqnarray}
	Z[{\overline{A}}]_{0} & = & 
	\exp\Biggl[-\int\!\!d^{4}x_{_{E}}\biggl\{{m^{2}\over2g^{2}}
	 +{1\over8\pi^{2}}\biggl[{\Lambda^{4}\over2}
	 \Bigl(1-{m^{2}\over\Lambda^{2}}\Bigr)
	 {\mathrm{e}}^{-{m^{2}\over\Lambda^{2}}}
	 +{m^{4}\over2}E_{1}({m^{2}/\Lambda^{2}})
	\nonumber  \\
	 &  & {}+{1\over3}\Bigl({1\over{\overline{\epsilon}}}
	+\ln{\mu^{2}\over m^{2}}\Bigr)
	{F_{\mu\nu}F_{\mu\nu}\over2}
	-{1\over45m^{4}}\biggl[
	\left({F_{\mu\nu}F_{\mu\nu}\over2}\right)^{2}
	-7\Bigl({F_{\mu\nu}{\widetilde{F}}_{\mu\nu}\over4}\Bigr)^{2}
	\biggr]\hspace*{20pt}
	\nonumber  \\
	 &  & 
	{}+{2\over315m^{8}}\biggl[
	2\left({F_{\mu\nu}F_{\mu\nu}\over2}\right)^{3}
	-13\Bigl({F_{\mu\nu}{\widetilde{F}}_{\mu\nu}\over4}\Bigr)^{2}
	{F_{\mu\nu}F_{\mu\nu}\over2}\biggr]+O(F^{8})
	\biggr]\biggr\}\Biggr],
	\label{ZA4}
\end{eqnarray}
where
\begin{eqnarray}
	E_{1}(z) & = & \int_{1}^{\infty}\!\!dt{{\mathrm{e}}^{-zt}\over t},
	\label{E1}  \\
	{1\over{\overline{\epsilon}}} & = & {1\over\epsilon}-\gamma+\ln2\pi,
	\label{epsilonb}
\end{eqnarray}
and $\mu$ is a renormalization scale. [Note that $E_{1}(z)>0$ for any 
real $z(>0)$.] Recall that the gauge action 
is written as $(-{1/4e_{bare}^{2}})F_{\mu\nu}F_{\mu\nu}$ to 
define a renormalized charge such that 
\begin{equation}
	{1\over e_{_{\mathrm{R}}}^{2}(\mu)}
	={Z_{3}\over e_{_{\mathrm{R}}}^{2}(\mu)}
	+{1\over12\pi^{2}}{1\over{\overline{\epsilon}}},
\end{equation}
where the first term of the right-hand side is the bare part and 
$Z_{3}$ is the wave function renormalization constant. 

Now we turn our attention to the functional integration of the gauge 
field strength. 
The total partition function is given by combining Eq.~(\ref{ZA4}) 
with the result in the preceding section: 
\begin{eqnarray}
Z[m]&=&\exp\left[-\int\!\!d^{4}x_{_{E}}V_{0}(m)\right]
\int\!\!{\mathcal{D}}{\beta}_{\alpha}{\mathcal{D}}\rho
[{\mathrm{Det}}\vert{\boldsymbol{\nabla}}\vert]^{-2}
\label{ZT}\\
&&{}\times\exp\left[-{1\over8\pi^{2}}\int\!\!d^{4}x_{_{E}}\biggl\{
\Bigl({4\pi^{2}\over e_{_{\mathrm{R}}}^{2}(\mu)}
+{1\over3}\ln{\mu^{2}\over m^{2}}\Bigr)
{{\overline{F}}_{\mu\nu}{\overline{F}}_{\mu\nu}\over2}
+\mbox{higher orders}
	\biggr\}\right], 
	\nonumber
\end{eqnarray}
where
\begin{equation}
	 V_{0}(m)={m^{2}\over2g^{2}}
	 +{1\over8\pi^{2}}\left[{\Lambda^{4}\over2}
	 \Bigl(1-{m^{2}\over\Lambda^{2}}\Bigr)
	 {\mathrm{e}}^{-{m^{2}\over\Lambda^{2}}}
	 +{m^{4}\over2}E_{1}({m^{2}/\Lambda^{2}})\right],
	 \label{vo}
\end{equation}
``higher orders'' are those of Eq.~(\ref{ZA4}) with 
$F_{\mu\nu}\rightarrow{\overline{F}}_{\mu\nu}$ , and 
${\overline{F}}_{\mu\nu}$ are given by Eqs.~(\ref{Fi4B}) and 
(\ref{FijB}). All variables in Eq.~(\ref{ZT}) are 
${\overline{F}}_{\mu\nu}{\overline{F}}_{\mu\nu}$ and 
${\overline{F}}_{\mu\nu}{\widetilde{\overline{F}}}_{\mu\nu}$, 
so we can change the integration variables from $\beta_{\alpha}$ and 
$\rho$ to $S$, $T$, and $U$: 
\begin{eqnarray}
	{\overline{F}}_{\mu\nu}{\overline{F}}_{\mu\nu}
	& = & 2(S^{2}+T^{2}+U^{2}),
	\label{CVFF}  \\
	{\overline{F}}_{\mu\nu}{\widetilde{\overline{F}}}_{\mu\nu} 
	& = & 2(S^{2}-T^{2}),
	\label{CVFFT}
\end{eqnarray}
yielding 
\begin{equation}
	{\mathcal{D}}{\beta}_{\alpha}{\mathcal{D}}\rho
	[{\mathrm{Det}}\vert{\boldsymbol{\nabla}}\vert]^{-2}
	={\mathcal{D}}S\,{\mathcal{D}}T\,{\mathcal{D}}U\,
	{\mathrm{Det}}[-\framebox(7,7){}_{_{E}}]^{-1},
\end{equation}
from Eq.~(\ref{FF}). 
As for ${\overline{F}}_{\mu\nu}{\widetilde{\overline{F}}}_{\mu\nu}$, 
though it cannot be expressed by the total divergence any more, its 
expectation value in the space-time integral would vanish,
\begin{equation}
	\int\!\!d^{4}x_{_{E}}\langle{\overline{F}}_{\mu\nu}
	{\widetilde{\overline{F}}}_{\mu\nu}\rangle
	=\int\!\!d^{4}x_{_{E}}\langle2(S^{2}-T^{2})\rangle=0,
\end{equation}
due to a symmetry $S\leftrightarrow T$ of the effective action 
(\ref{ZT}); the dependence on $S$ and $T$ appears only in the forms 
$S^{2}+T^{2}$ and $(S^{2}-T^{2})^{2}$. 

Note that $Z[m]$ (\ref{ZT}) is a trivial product of integrals at each 
space-time point. We evaluate these integrals with the help of the WKB 
approximation. For a technical reason we discard $O(F^{8})$ terms in 
Eq.~(\ref{ZA4}). (This enables us to obtain the stationary point 
analytically. See below.) 
Discretizing space-time with 
$(\mbox{lattice spacing})^{4}=32\pi^{2}/\Lambda^{4}$ and neglecting 
the irrelevant factor ${\mathrm{Det}}[-\framebox(7,7){}_{_{E}}]^{-1}$, 
we obtain
\begin{eqnarray}
	Z[x]&=&\prod_{\{\mbox{site}\}}8\,(32\pi)^{{3\over2}}
	\int_{[0,\infty)^{3}}\!\!ds\,dt\,du\,
	\nonumber  \\
	&&{}\times\exp\Biggl[-\biggl\{v_{0}(x)
	+\Bigl({16\pi^{2}\over e_{_{\mathrm{R}}}^{2}(\mu)}
	+{4\over3}\ln{\mu^{2}\over x\Lambda^{2}}\Bigr)(s^{2}+t^{2}+u^{2})
	\nonumber\\
	&&{}-{1\over45x^{2}}
	\left[4(s^{2}+t^{2}+u^{2})^{2}-7(s^{2}-t^{2})^{2}\right]
	\nonumber  \\
	&&{}+{2\over315x^{4}}\left[8(s^{2}+t^{2}+u^{2})^{3}
	-13(s^{2}+t^{2}+u^{2})(s^{2}-t^{2})^{2}\right]\biggr\}\Biggr],
	\label{ZT1}
\end{eqnarray}
where the first coefficients are normalization factors to cancel the 
Gaussian integration; 
dimensionless parameters, $x=m^{2}/\Lambda^{2}$, 
$s=S/\Lambda^{2}$, etc. 
have been introduced; and 
\begin{equation}
	v_{0}(x)\equiv{32\pi^{2}V_{0}(m)\over\Lambda^{4}}
	={16\pi^{2}x\over g^{2}\Lambda^{2}}
	 +2(1-x){\mathrm{e}}^{-x}
	 +2x^{2}E_{1}(x).
\end{equation}
Introducing a polar coordinate $(r,\theta,\varphi)$ with 
$r^{2}=s^{2}+t^{2}+u^{2}$ and 
$r^{2}\sin^{2}\theta\cos2\varphi=s^{2}-t^{2}$ and integrating with 
respect to $\theta$ and $\varphi$, we find 
\begin{equation}
	Z[x]=\prod_{\{\mbox{site}\}}4\pi\,(32\pi)^{{3\over2}}
	\int_{0}^{\infty}\!\!dr\,\exp\left[-\left\{v_{0}(x)
	+v_{_{F}}(r;x)\right\}\right], 
	\label{ZT2}
\end{equation}
where 
\begin{eqnarray}
	v_{_{F}}(r;x) & = & A(x)r^{2}-B(x)r^{4}+C(x)r^{6}-\ln r^{2},
	\label{vf}  \\
	A(x) & = & {16\pi^{2}\over e_{_{\mathrm{R}}}^{2}(\mu)}
	+{4\over3}\ln{\mu^{2}\over x\Lambda^{2}}=
	{16\pi^{2}\over e_{_{\mathrm{R}}}^{2}(\Lambda)}
	-{4\over3}\ln{x},
	\label{ax}  \\
	B(x) & = & {32\over675x^{2}},
	\label{bx}\\
	C(x) & = & {136\over4725x^{4}},
	\label{cx}
\end{eqnarray}
and $O(r^{8})$ terms in the exponent has been neglected. 

\begin{figure}[bt]
	\begin{center}
		\scalebox{0.65}{\includegraphics{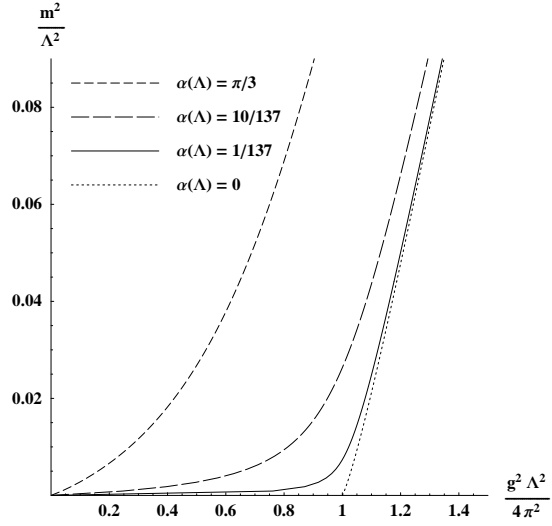}}
	\end{center}
	\caption{Squared dynamical mass of fermion $m^{2}/\Lambda^{2}$ shown 
	as a function of the four-fermion coupling constant 
	$g^{2}\Lambda^{2}/4\pi^{2}=G$ for several fixed gauge coupling 
	constants 
	$\alpha(\Lambda)={\mathrm{e}}_{_{\mathrm{R}}}^{2}(\Lambda)/4\pi$, 
	obtained from Eq.~(\ref{Gap0}).}
	\label{Fig1}
\end{figure}

Let us first consider the lowest correction, up to the $O(r^{2})$ term. 
The $r$-integration is analytically performed to give 
\begin{equation}
	Z[x]=\prod_{\{\mbox{site}\}}2^{15/2}\pi^{3}\exp[-v(x)]
\end{equation}
where 
\begin{eqnarray}
	v(x)&=&v_{0}(x)+{3\over2}\ln A(x)
	\nonumber\\
	&=&{4x\over G}+2(1-x){\mathrm{e}}^{-x}+2x^{2}E_{1}(x)
	+{3\over2}\ln\left[
	{4\pi\over\alpha(\Lambda)}-{4\over3}\ln x\right]
\end{eqnarray}
with $G=g^{2}\Lambda^{2}/4\pi^{2}$ and 
$\alpha(\Lambda)=e_{_{\mathrm{R}}}^{2}(\Lambda)/4\pi$. 
The argument of the logarithm becomes negative at the Landau pole 
$x=\exp[3\pi/\alpha(\Lambda)]$, but we do not care such a heavy 
fermion and thus assume that the argument is always positive. 
The stationary point $x^{*}$ defined by 
\begin{equation}
	{1\over4}{\partial v(x)\over\partial x}\Bigg\vert_{x=x^{*}}
	={1\over G}-{\mathrm{e}}^{-x^{*}}+x^{*}E_{1}(x^{*})
	-{3\alpha(\Lambda)\over8(3\pi-\alpha(\Lambda)\ln x^{*})x^{*}}=0
	\label{Gap0}
\end{equation}
should satisfy the stability condition 
\begin{equation}
	{1\over4}{\partial^{2}v(x)\over\partial x^{2}}\Bigg\vert_{x=x^{*}}
	=E_{1}(x^{*})+{9\pi\alpha(\Lambda)-3\alpha(\Lambda)^{2}
	(\ln x^{*}+1)\over8(3\pi-\alpha(\Lambda)\ln x^{*})^{2}x^{*2}}>0,
\end{equation}
for which $x^{*}<{\mathrm{e}}^{-1}\simeq0.368$ is a sufficient 
condition. To solve 
Eq.~(\ref{Gap0}) we set a condition:
\begin{equation}
	1\gg-x\ln x\gg(1-\gamma)x\gg x^{2},
	\label{genecon}
\end{equation}
which is fulfilled if $x<1.\times10^{-2}$. (In the actual 
situation $m\sim1$ MeV and we should take 
$\Lambda>1$ TeV, a lower bound of compositeness from 
experiments\cite{PDG}, to give $x<10^{-12}$.) Under this condition 
Eq.~(\ref{Gap0}) becomes 
\begin{equation}
	{1\over G}-1-x\ln x
	-{3\alpha(\Lambda)\over8(3\pi-\alpha(\Lambda)\ln x)x}=0,
	\label{Gap1}
\end{equation}
where we have used the expansion of $E_{1}(x)$ for $x\ll1$, 
\begin{equation}
	E_{1}(x)=-\gamma-\ln x+O(x).
\end{equation}
For $\alpha(\Lambda)=0$ there exists a nonvanishing solution only if 
$G\geq1$; therefore the critical coupling $G_{c}$ is 1. 
The solution for $G\simeq1$ is 
\begin{equation}
	x^{*}\simeq-{G-1\over\ln[G-1]}.
	\label{sol1}
\end{equation}
For $\alpha(\Lambda)>0$ the solution is obtained in two separate regions: 
(i) $x\ll\exp\left[-{3\pi\over\alpha(\Lambda)}\right]$; 
\begin{equation}
	x^{*}\simeq-{3G\over8\ln({3G\over8})},
	\label{sol2}
\end{equation}
which is independent of $\alpha(\Lambda)$. (ii) 
$\exp\left[-{3\pi\over\alpha(\Lambda)}\right]\ll x<\alpha(\Lambda)/4$; 
\begin{equation}
	x^{*}\simeq{\alpha(\Lambda)G\over8\pi(1-G)}.
	\label{sol3}
\end{equation}
Thus there always exists a nonvanishing solution for a given $G$, 
that is, $G_{c}=0$. [Solutions at several $\alpha(\Lambda)$'s 
are depicted in Fig.~\ref{Fig1}.] The actual situation, 
$\alpha(\Lambda)\simeq1/137$, $x\simeq10^{-12}$, and 
$\Lambda\simeq1$ TeV, lies in the case (ii) 
($\exp\left[-{3\pi\over\alpha(\Lambda)}\right]\simeq1.7\times10^{-561}$) 
and from Eq.~(\ref{sol3}) 
\begin{equation}
	\pi G={g^{2}\Lambda^{2}\over4\pi}\simeq{8\pi^{2}x\over\alpha(\Lambda)}
	\simeq1.1\times10^{-8};
\end{equation}
$g^{2}$ is highly suppressed even at this scale 
($\Lambda\simeq1$ TeV). 
%
%

Now we include the higher order terms. We evaluate the 
$r$-integration with the WKB approximation; 
\begin{equation}
	\int_{0}^{\infty}\!\!dr\,\exp\left[-v_{_{F}}(r;x)\right]
	\simeq(2\pi)^{{1\over2}}\exp\left[-\left\{v_{_{F}}(r_{0}(x);x)
	+{1\over2}\ln\left[v''_{_{F}}(r_{0}(x);x)\right]\right\}\right], 
\end{equation}
where the superscript ${}'$ denotes the $r$-differentiation 
and $r_{0}(x)$ is a solution of the stationary point equation 
\begin{equation}
	v'_{_{F}}(r;x)={2\over r}
	\left[A(x)r^{2}-2B(x)r^{4}+3C(x)r^{6}-1\right]=0,
	\label{stpeq}
\end{equation}
which is the cubic equation of $r^{2}$. 
There exists only one real-positive solution for $r^{2}$, 
\begin{equation}
	r_{0}^{2}(x)={56\over153}x^{2}
	+{3\over2}\left({175\over102}\right)^{{1\over3}}x^{{4\over3}}
	\left\{\left[P(x)+Q(x)\right]^{{1\over3}}
	-\left[P(x)-Q(x)\right]^{{1\over3}}\right\},
\end{equation}
where 
\begin{eqnarray}
	P(x) & = & \sqrt{1+{802816\over47403225}x^{2}
	 -{112\over153}x^{2}A(x)-{3136\over70227}x^{4}A(x)^{2}
	 +{175\over102}x^{4}A(x)^{3}},\hspace*{30pt}
	\label{Px}  \\
	Q(x) & = & 1+{401408\over47403225}x^{2}
	 -{56\over153}x^{2}A(x),
	\label{Qx}
\end{eqnarray}
since $C(x)>0$ from (\ref{cx}) and $9A(x)C(x)-4B(x)^{2}>0$ turns out 
to be 
\begin{equation}
	\alpha(\Lambda)<{103275\pi\over896}
	\simeq3.62\times10^{2}.
	\label{condition}
\end{equation}
Positivity of $v''_{_{F}}(r_{0}(x);x)$ 
\begin{equation}
	v''_{_{F}}(r;x)=2\left[15C(x)r^{4}-6B(x)r^{2}+A(x)
	+{1\over r^{2}}\right]>0
\end{equation}
is also guaranteed, since 
the sufficient condition $9B(x)^{2}-15A(x)C(x)<0$ 
is fulfilled. [This leads to a similar condition as 
Eq.~(\ref{condition}).]

\begin{figure}[bt]
	\begin{center}
		\scalebox{0.65}{\includegraphics{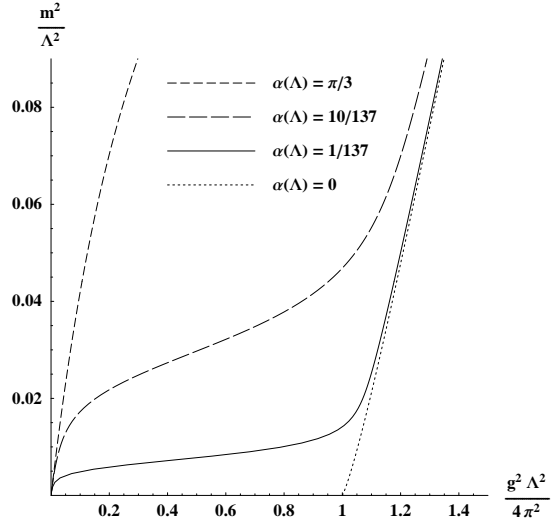}}
	\end{center}
	\caption{Squared dynamical mass of fermion $m^{2}/\Lambda^{2}$ shown 
	as a function of the four-fermion coupling constant 
	$g^{2}\Lambda^{2}/4\pi^{2}=G$ for several fixed gauge coupling 
	constants 
	$\alpha(\Lambda)={\mathrm{e}}_{_{\mathrm{R}}}^{2}(\Lambda)/4\pi$, 
	obtained from Eq.~(\ref{vfinal}).}
	\label{Fig2}
\end{figure}

The final form of the partition function is 
\begin{equation}
	Z[x]=\prod_{\{\mbox{site}\}}2^{10}\pi^{3}\exp[-v(x)],
\end{equation}
with 
\begin{eqnarray}
	v(x) & = & v_{0}(x)+v_{_{F}}(r_{0}(x);x)
	+{1\over2}\ln\left[v''_{_{F}}(r_{0}(x);x)\right]
	\nonumber  \\
	 & = & v_{0}(x)+{1\over3}+{2\over3}A(x)r_{0}(x)^{2}
	 -{1\over3}B(x)r_{0}(x)^{4}
	\label{vfinal}  \\
	 &  & {}+{1\over2}\ln
	 \left[{12C(x)\{3-2A(x)r_{0}(x)^{2}+2B(x)r_{0}(x)^{4}\}
	 \over1-A(x)r_{0}(x)^{2}+2Br_{0}(x)^{4}}\right],
	\nonumber
\end{eqnarray}
where Eq.~(\ref{stpeq}) has been used. Differentiating this with 
respect to $x$ 
gives us the gap equation, whose numerical solution is 
shown in Fig.~\ref{Fig2}. In comparison with Fig.~\ref{Fig1}, 
there is a large deviation in the region where 
both $x=m^{2}/\Lambda^{2}$ and $G=g^{2}\Lambda^{2}/4\pi^{2}$ are small, 
and the mass is increased by higher order 
corrections.  
%
%

For sufficiently small $x$, $xA(x)\ll1$ or 
$x\ll\alpha(\Lambda)/4\pi$, Eq.~(\ref{vfinal}) can be expanded: 
\begin{eqnarray}
	v(x) & = & v_{0}(x)+{1\over3}
	-{1\over2}\ln{525x^{4}\over544}
	-{8\over3}\left({7\over255}\right)^{2/3}x^{2/3}
	\nonumber  \\
	&&{}+{25\over3}\left({7\over255}\right)^{1/3}x^{4/3}
	\left({3\pi\over\alpha(\Lambda)}
	-{1792\over172125}-\ln x\right)
	\label{vfinale}  \\
	&&{}+{448\over1377}x^{2}\left({3\pi\over\alpha(\Lambda)}
	-{224\over20655}-\ln x\right)
	+O(x^{8/3}),
	\nonumber
\end{eqnarray}
yielding the gap equation 
\begin{eqnarray}
	{1\over4}{\partial v(x)\over\partial x}
	&=&{1\over G}-{\mathrm{e}}^{-x}
	+xE_{1}(x)
	-{1\over2x}
	-\left({7\over255}\right)^{2/3}{4\over9x^{1/3}}
	\nonumber\\
	&&{}+{25\over9}\left({7\over255}\right)^{1/3}x^{1/3}
	\left({3\pi\over\alpha(\Lambda)}
	-{523543\over688500}-\ln x\right)
	\label{gape}  \\
	&&{}+{224\over1377}x\left({3\pi\over\alpha(\Lambda)}
	-{21103\over41310}-\ln x\right)+O(x^{5/3}).
	\nonumber
\end{eqnarray}
The stability condition 
\begin{eqnarray}
	{1\over4}{\partial^{2} v(x)\over\partial x^{2}}
	&=&E_{1}(x)
	+{1\over2x^{2}}
	+\left({7\over255}\right)^{2/3}{4\over27x^{4/3}}
	\nonumber\\
	&&{}+\left({7\over255}\right)^{1/3}{25\over27x^{2/3}}
	\left({3\pi\over\alpha(\Lambda)}
	-{2589043\over688500}-\ln x\right)
	\label{state}  \\
	&&{}
	+{224\over1377}\left({3\pi\over\alpha(\Lambda)}
	-{62413\over41310}-\ln x\right)+O(x^{2/3})>0, 
	\nonumber
\end{eqnarray}
is fulfilled for $x<1$ and 
$\alpha(\Lambda)<{2065500\over2589043}\pi$. Therefore an 
$\alpha$-independent solution 
exists: 
\begin{equation}
	x^{*}\simeq{G\over2(1-G)}
\end{equation}
for $x$ obeying $x\ll\alpha(\Lambda)/4\pi$ and Eq.~(\ref{genecon}). 
In this case $g^{2}$ is also highly suppressed:
\begin{equation}
	\pi G={g^{2}\Lambda^{2}\over4\pi}\simeq2x\simeq2.\times10^{-12},
\end{equation}
for $x\simeq10^{-12}$, that is, $\Lambda\simeq1$ TeV.

\section{Discussion}\label{DISC}
We give a gauge invariant 
recipe for calculating the effective action in QED with the four-fermion 
interaction. We use perturbation and the local potential 
approximation to study the dynamical fermion mass. 
In order to include the higher orders of the gauge coupling constant 
we just expand the completely 
known function (\ref{G}) into a polynomial of 
$F_{+}^{2}+F_{-}^{2}[=F_{\mu\nu}F_{\mu\nu}/2]$ and 
$F_{+}^{2}F_{-}^{2}[=(F_{\mu\nu}\widetilde{F}_{\mu\nu}/4)^{2}]$ and 
integrate with respect to the proper-time $\tau$. The last task is to 
evaluate the triple integral of $s$, $t$, and $u$. In this 
article we employ the WKB approximation for it, which, however, 
cannot be performed in an elementary manner when higher orders are 
included further. Meanwhile the ordinary perturbative treatment, 
considering the Gaussian part the kernel and 
expanding the exponential of higher parts than the third order, can 
always be ensured. 


As for the local potential approximation its 
efficacy is still open. In order to go beyond the local potential 
approximation, we must perform the functional integration of $S$, 
$T$, and $U$, instead of the ordinary integration like (\ref{ZT1}). 
For example, an $O(\alpha(\Lambda))$ 
quantity after integrating with respect to the gauge fields would be 
\begin{eqnarray}
	I&=&\int\!\!d^{^{D}}\!x_{_{E}}\,d^{^{D}}\!y_{_{E}}\,\langle J_{\mu}
	A_{\mu}(x)J_{\rho}A_{\rho}(y)\rangle
	\bigg/\int\!\!d^{^{D}}\!x_{_{E}}
	\nonumber\\
	&=&\int\!\!d^{^{D}}\!x_{_{E}}\,d^{^{D}}\!y_{_{E}}\,\langle J_{\mu}
	{\partial_{\nu}F_{\mu\nu}\over\framebox(7,7){}_{_{E}}}(x)
	J_{\rho}{\partial_{\sigma}F_{\rho\sigma}\over
	\framebox(7,7){}_{_{E}}}(y)\rangle
	\bigg/\int\!\!d^{^{D}}\!x_{_{E}}
	\nonumber\\
	&=&\int\!\!{d^{^{D}}\!p_{_{E}}\over(2\pi)^{^{D}}}\Pi(p^{2})
	\left(p^{2}\delta_{\mu\rho}-p_{\mu}p_{\rho}\right)
	{p_{\nu}p_{\sigma}\over(p^{2})^{2}}
	\langle{F}_{\mu\nu}(p)
	{F}_{\rho\sigma}(-p)\rangle,
\end{eqnarray}
where 
\begin{equation}
	\langle{F}_{\mu\nu}(p)
	{F}_{\rho\sigma}(-p)\rangle=
	{1\over p^{2}}
	\left(p_{\mu}p_{\rho}\delta_{\nu\sigma}
	-p_{\mu}p_{\sigma}\delta_{\nu\rho}
	-p_{\nu}p_{\rho}\delta_{\mu\sigma}
	+p_{\nu}p_{\sigma}\delta_{\mu\rho}\right),
\end{equation}
so that 
\begin{equation}
	I=\int\!\!{d^{^{D}}\!p_{_{E}}\over(2\pi)^{^{D}}}\Pi(p^{2})(D-1)
	=\int\!\!{d^{^{D}}\!p_{_{E}}\over(2\pi)^{^{D}}}\Pi(p^{2})
	{\langle{F}_{\mu\nu}(p)
	{F}_{\mu\nu}(-p)\rangle\over2}.
	\label{I}
\end{equation}
The local potential approximation is to expand $\Pi(p^{2})$ in terms 
of $p^{2}$ and keep the lowest. Since the change of variables, 
Eq.~(\ref{CVFF}), gives 
${F}_{\mu\nu}(p){F}_{\mu\nu}(-p)=S(p)S(-p)+T(p)T(-p)+U(p)U(-p)$, 
the leading correction to the local potential approximation is 
\begin{equation}
	\int\!\!d^{D}x\,\Pi'(0)(-S\,\framebox(7,7){}_{_{E}}S
	-T\,\framebox(7,7){}_{_{E}}T
	-U\,\framebox(7,7){}_{_{E}}U).
\end{equation}
Therefore we must calculate the effective potential with the help of 
the Feynman graphs using the propagators of $S$, $T$, and $U$. 
%

Our result indicates that 
chiral symmetry is always broken for any $\alpha(\Lambda)$ and $G$ as 
is seen from Figs.~\ref{Fig1} and \ref{Fig2}.
This seems to be different from previous 
results\cite{KMY,ASTW,AMSTT,Ko} which claim that there is a nonzero 
$G_{c}$ if $\alpha(\Lambda)<\pi/3$. It is, however, too early to 
conclude, since our present work is restricted only to the tree level of 
auxiliary fields 
and thus no higher orders of $G$ are included.  
As is seen from the Figs.~\ref{Fig1} and \ref{Fig2}, the small 
negative correction, linear in $G$, to dynamical mass could easily 
swallow the broken region. From a recent study\cite{Ka}, the one-loop 
inclusion of the auxiliary fields is promising. 
Thus a further study must be necessary for a definite conclusion. 

In QED the functional integral in terms of the field strength can be 
constructed, since the configuration space of the gauge 
potential as well as the field strength is trivial enough for 
the gauge to be completely fixed. 
It is challenging to generalize Eq.~(\ref{ZF}) to QCD where an obstacle for 
gauge fixing, the Gribov ambiguity\cite{Gribov}, exists. 
In order to examine the 
dynamics of chiral symmetry breaking and color confinement in QCD, this 
must be done and the choice of convenient variables describing the low 
energy phenomena\cite{FN} must be necessary. 

These directions of study are in progress.  

\section*{Acknowledgments}
The author thanks T.~Kashiwa and K.~Harada for discussions and encouragements 
and also thanks D.~McMullan for discussions. 

\end{document}